\documentclass[twocolumn,longauthor]{aastex63}
\usepackage{hyperref}
\usepackage{soul}

\def \src {\mbox{SGR~1935$+$2154}}
\def \frbdic {\mbox{FRB~180916.J0158$+$65}}
\def \xmm {\emph{XMM-Newton}}

\def \pdot {\dot P}

\graphicspath{{fig/}}

\defcitealias{Davies81}{DP81} %usage: \citetalias{Davies81}

\received{May 12, 2020 }
\revised{ June 10, 2020}
\accepted{July 3, 2020 }
\submitjournal{ApJ Letters}

\shorttitle{\src}
\shortauthors{Mereghetti et al.}
\NewPageAfterKeywords

\begin{document}

\title{INTEGRAL discovery of a burst with associated radio emission from the magnetar \src\
% \footnote{Based on observations  obtained with \xmm, an ESA science  mission  with  instruments  and  contributions  directly  funded  by ESA Member States and NASA.}
}

\correspondingauthor{Sandro Mereghetti}
\email{sandro.mereghetti@inaf.it}

\author[0000-0003-3259-7801]{S.~Mereghetti} %$^{1}$
\affiliation{INAF -- Istituto di Astrofisica Spaziale e Fisica Cosmica, Via A. Corti 12, I-20133 Milano, Italy}

\author[0000-0001-6353-0808]{V.~Savchenko} %$^{2}$,
\affiliation{ISDC, Department of Astronomy, University of Geneva, Chemin d'\'Ecogia, 16 CH-1290 Versoix, Switzerland}

\author[0000-0003-1429-1059]{C.~Ferrigno} %$^{2}$,
\affiliation{ISDC, Department of Astronomy, University of Geneva, Chemin d'\'Ecogia, 16 CH-1290 Versoix, Switzerland}

\author[0000-0001-9494-0981]{D.~G\"otz} %$^{3}$,
\affiliation{AIM-CEA/DRF/Irfu/D\'epartement d’Astrophysique, CNRS, Universit\'e Paris-Saclay, Universit\'e de Paris, \\
	\hspace{0.05cm}Orme des Merisiers, F-91191 Gif-sur-Yvette, France}

\author[0000-0001-6641-5450]{M.~Rigoselli} %$^{1}$
\affiliation{INAF -- Istituto di Astrofisica Spaziale e Fisica Cosmica, Via A. Corti 12, I-20133 Milano, Italy}

\author[0000-0002-6038-1090]{A.~Tiengo} %$^{1}$
\affiliation{Scuola Universitaria Superiore IUSS, Piazza della Vittoria 15, I-27100 Pavia, Italy}

\author[0000-0002-2017-4396]{A.~Bazzano} %$^{4}$
\affiliation{INAF -- Institute for Space Astrophysics and Planetology, Via Fosso del Cavaliere 100, I-00133 Rome, Italy}

\author{E.~Bozzo} %$^{2}$,
\affiliation{ISDC, Department of Astronomy, University of Geneva, Chemin d'\'Ecogia, 16 CH-1290 Versoix, Switzerland}

\author[0000-0003-0860-440X]{A.~Coleiro} %$^{5}$,
\affiliation{APC, AstroParticule et Cosmologie, Universit\'e Paris Diderot, CNRS/IN2P3, CEA/Irfu, Observatoire de Paris Sorbonne Paris Cit\'e,\\
	\hspace{0.05cm}10 rue Alice Domont et L\'eonie Duquet, F-75205 Paris Cedex 13, France.}

\author[0000-0003-2396-6249]{T.~J.-L.~Courvoisier} %$^{2}$,
\affiliation{ISDC, Department of Astronomy, University of Geneva, Chemin d'\'Ecogia, 16 CH-1290 Versoix, Switzerland}

\author[0000-0001-8538-4864]{M.~Doyle} %$^{7}$,
\affiliation{Space Science Group, School of Physics, University College Dublin, Belfield, Dublin 4, Ireland}

\author{A.~Goldwurm} %$^{5}$,
\affiliation{APC, AstroParticule et Cosmologie, Universit\'e Paris Diderot, CNRS/IN2P3, CEA/Irfu, Observatoire de Paris Sorbonne Paris Cit\'e,\\
	\hspace{0.05cm}10 rue Alice Domont et L\'eonie Duquet, F-75205 Paris Cedex 13, France.}

\author[0000-0003-2931-3732]{L.~Hanlon} %$^{7}$,
\affiliation{Space Science Group, School of Physics, University College Dublin, Belfield, Dublin 4, Ireland}

\author[0000-0001-9932-3288]{E.~Jourdain} %$^{8}$
\affiliation{CNRS; IRAP; 9 Av. colonel Roche, BP 44346, F-31028 Toulouse cedex 4, France}
\affiliation{Universit\'e de Toulouse; UPS-OMP; IRAP;  Toulouse, France}

\author[0000-0002-0221-5916]{A.~von~Kienlin} %$^{6}$,
\affiliation{Max-Planck-Institut f\"{u}r Extraterrestrische Physik, Garching, Germany}

\author{A.~Lutovinov} %$^{9,10}$,
\affiliation{Space Research Institute of Russian Academy of Sciences, Profsoyuznaya 84/32, 117997 Moscow, Russia}

\author[0000-0001-5108-0627]{A.~Martin-Carrillo} %$^{7}$,
\affiliation{Space Science Group, School of Physics, University College Dublin, Belfield, Dublin 4, Ireland}

\author{S.~Molkov} %$^{9,10}$,
\affiliation{Space Research Institute of Russian Academy of Sciences, Profsoyuznaya 84/32, 117997 Moscow, Russia}

\author[0000-0002-6601-9543]{L. Natalucci} %$^{4}$,
\affiliation{INAF -- Institute for Space Astrophysics and Planetology, Via Fosso del Cavaliere 100, I-00133 Rome, Italy}

\author[0000-0001-6286-1744]{F.~Onori} %$^{4}$,
\affiliation{INAF -- Institute for Space Astrophysics and Planetology, Via Fosso del Cavaliere 100, I-00133 Rome, Italy}

\author[0000-0003-0543-3617]{F.~Panessa} %$^{4}$,
\affiliation{INAF -- Institute for Space Astrophysics and Planetology, Via Fosso del Cavaliere 100, I-00133 Rome, Italy}

\author[0000-0003-2126-5908]{J.~Rodi} %$^{4}$,
\affiliation{INAF -- Institute for Space Astrophysics and Planetology, Via Fosso del Cavaliere 100, I-00133 Rome, Italy}

\author[0000-0002-4151-4468]{J.~Rodriguez} %$^{3}$,
\affiliation{AIM-CEA/DRF/Irfu/D\'epartement d’Astrophysique, CNRS, Universit\'e Paris-Saclay, Universit\'e de Paris, \\
	\hspace{0.05cm}Orme des Merisiers, F-91191 Gif-sur-Yvette, France}

\author{C.~S\'anchez-Fern\'andez} %$^{11}$
\affiliation{European Space Astronomy Centre (ESA/ESAC), Science Operations Department, 28691 Villanueva de la Ca\~nada, Madrid, Spain}

\author{R.~Sunyaev} %$^{11,12}$,
\affiliation{Space Research Institute of Russian Academy of Sciences, Profsoyuznaya 84/32, 117997 Moscow, Russia}
\affiliation{Max Planck Institute for Astrophysics, Karl-Schwarzschild-Str. 1, Garching b. D-85741, Munchen, Germany}

\author[0000-0003-0601-0261]{and P.~Ubertini} %$^{4}$
\affiliation{INAF -- Institute for Space Astrophysics and Planetology, Via Fosso del Cavaliere 100, I-00133 Rome, Italy}

\begin{abstract}
We report on INTEGRAL observations of the soft $\gamma$-ray repeater \src\      performed between 2020 April 28 and May 3.  Several short bursts with fluence of $\sim10^{-7}-10^{-6}$  erg cm$^{-2}$ were detected by the IBIS instrument in the 20-200 keV range.  The burst with the hardest spectrum, discovered and localized in real time by the INTEGRAL Burst Alert System,  was spatially  and temporally coincident with a short and very bright radio burst 
detected by the CHIME and STARE2 radio telescopes at 400-800 MHz and 1.4 GHz, respectively.
Its lightcurve  shows three narrow peaks
separated by  $\sim$29 ms time intervals, superimposed on a broad pulse lasting
$\sim$0.6 s. The brightest peak had a  delay of 6.5$\pm$1.0 ms with respect to the 1.4 GHz radio pulse (that coincides with the second and brightest component seen at lower frequencies). 
The burst spectrum,   an exponentially cut-off power law with photon index $\Gamma=0.7_{-0.2}^{+0.4}$ and peak energy $E_p=65\pm5$ keV, is harder than those of the  bursts usually observed from this and other magnetars.
By the analysis  of an expanding dust scattering ring seen in X-rays with the {\it Neil Gehrels Swift Observatory} XRT instrument, we  derived a distance of 4.4$_{-1.3}^{+2.8}$ kpc  for  \src , independent of its possible association with the supernova remnant G57.2+0.8. At this distance, the burst 20-200 keV fluence of
  $(6.1\pm 0.3)\times10^{-7}$ erg cm$^{-2}$ corresponds to an isotropic emitted energy of $\sim1.4\times10^{39}$ erg.
This is the first burst with a radio counterpart observed from a soft $\gamma$-ray  repeater  and it strongly supports     models based on magnetars that have been proposed  for extragalactic fast radio bursts.
\end{abstract}



\section{Introduction}
\label{sec:intro}

Magnetars are neutron stars powered mainly by magnetic energy dissipation that are characterized by strong variability on several timescales   (see \citealt{mer15,tur15,kas17}  for reviews). Many known magnetars were first discovered  as Soft Gamma-Ray Repeaters (SGRs), i.e. sources of multiple bursts of hard X-rays, typically
shorter than 1 s and with peak  luminosity up to $\sim10^{39-41}$ erg s$^{-1}$.  Longer and more energetic bursts have also been observed from a few SGRs, with the most extreme cases, known as Giant Flares, reaching peak luminosity as large as 10$^{47}$ erg s$^{-1}$ and releasing a total energy up to 10$^{46}$  erg (e.g., \citet{pal05}). All the currently known magnetars are in the Galaxy or in the Magellanic Clouds \citep{ola14}.

Fast Radio Bursts (FRBs, see \citet{cor19,pet19} for reviews) are an enigmatic class of sources emitting short ($\sim1-10$ ms) pulses of radio emission with peak fluxes of $\sim0.1-100$ Jy at 1.4 GHz, and  dispersion measures (DM) in excess of the Milky Way values along their lines of sight.
Their extragalactic nature has been demonstrated by the association of a few FRBs with  host
galaxies
in the redshift range   0.0337-0.4755  \citep{marc20,pro19}, and  possibly up to  z=0.66 \citep{rav19}.
 A large variety of models have been proposed for FRBs (see \citealt{pla19} for an extensive and updated list), but no consensus has been reached yet on the nature of these sources.
Although many of these models involve  magnetars, there has been no confirmed association between an FRB and a high-energy bursting source until now.

Here we report on  hard X-ray  observations of the magnetar \src\
carried out with the INTEGRAL satellite in April--May 2020.  In particular, we present the properties of a short and hard burst  for which a spatially and temporally coincident FRB has been detected in the 400-800 MHz band \citep{arxiv-CHIME}
%\citep{atel13681}
and at 1.4 GHz \citep{arxiv-STARE}.
%\citep{atel13684}.
This is  the first detection of a pulse of radio emission clearly associated with an SGR burst, thus providing an intriguing connection between magnetars and FRBs.
We also report a new   estimate of the distance of \src\ based on the analysis of a dust scattering ring  detected in soft X-rays with the {\it Neil Gehrels SWIFT Observatory} \citep{atel13679}.

\section{\src\ } \label{sec:sgr}

\src\ was discovered through the detection of hard X-ray bursts with the Swift/BAT instrument in July 2014 \citep{sta14}. Its magnetar nature was confirmed by follow-up X-ray observations that revealed  a spin period $P$= 3.25 s and period derivative $\pdot = 1.43\times10^{-11}$ s s$^{-1}$ \citep{isr16},  corresponding to a characteristic age of 3.6 kyr and a dipole magnetic field of $2.2\times10^{14}$ G. A very faint (H$\sim$24) near infrared  counterpart was identified, thanks to its variability,   by the Hubble Space Telescope \citep{lev18}.
Pulsed radio emission has been observed in a
few transient magnetars, but not in \src\  \citep{bur14,sur14,isr16,atel13713}, with the most constraining upper limits of   14\,\mbox{$\mu$Jy} (4.6 GHz) and  7\,\mbox{$\mu$Jy} (1.4 GHz) derived by \citet{you17}. Single pulses were also undetected down to limits of $\sim$70 mJy  \citep{isr16} and $\sim$10-20 mJy  \citep{you17}.

Like most known magnetars, \src\ lies in the Galactic Plane  (l=57.25 deg, b=+0.82 deg), at an unknown distance.
The magnetar could be associated with the supernova remnant G57.2$+$0.8, for which  distances of $\sim$12.5 kpc  \citep{kot18}, 4.5--9 kpc  \citep{ran18}, and 6.6$\pm$0.7 \citep{zho20} have been derived. In the following, we will scale all the quantities to the distance   $d=4.4 d_{4.4}$ kpc, derived from our analysis of the dust scattering ring (see Section 4).

Since   its discovery, \src\ has been rather active, emitting short bursts in February 2015 and  May-July 2016 \citep{you17,lin20}, as well as an intermediate flare on April 12, 2015 \citep{koz16}.
Recently, \src\ has entered a new period of activity \citep{gcn27625,gcn27531} that culminated with the emission of  a ``burst forest'', i.e.  tens of bursts in  a short time interval, on  April 27-28 \citep{atel13675,atel13678,gcn27659}.

\begin{table}
        \caption{Log of INTEGRAL observations of \src\ in April-May 2020.
        %and number of bursts detected with IBIS/ISGRI.
        \label{tab-obslog}}
        \begin{center}
%        \begin{tabular}{lcccc}
        \begin{tabular}{lccc}
            \hline
   %                       Start time                &  End time   & ON-time\tablenotemark{a} & Off-axis & \# of \\
                          Start time                &  End time   & ON-time\tablenotemark{a} & Off-axis   \\
                                        \multicolumn{2}{c}{MM-DD HH:MM (UTC)}         & ks & angle($^o$)   \\
%                                        \multicolumn{2}{c}{MM-DD HH:MM (UTC)}         & ks & angle($^o$) & bursts \\
       \hline
       \hline

            04-22 18:54  & \
            04-23 07:44  & \
            41.1(89\%) & \
            6.2 -- 17.0  \
  %          0  \

            %& Compl.\tablenotemark{b} &
            %               2220:
            %                           %                    2 -- 25 & \
            %
               \\

            04-25 10:46  & \
            04-26 01:29  & \
            42.8(80\%) & \
            4.2 -- 18.6  \
 %           0  \

            %& Compl.\tablenotemark{b} &
            %               2221:
            %                           %                    2 -- 26 & \
            %
               \\

            04-28 02:37  & \
            04-29 12:54  & \
            102.6(83\%) & \
            0.5 -- 18.2  \
  %          8  \

            %& Compl.\tablenotemark{b} &
            %               2222:
            %                           %                    2 -- 61 & \
            %
               \\

            04-30 18:27  & \
            04-30 23:08  & \
            8.6(51\%) & \
            12.9 -- 16.0  \
   %         0  \

            %& Compl.\tablenotemark{b} &
            %               2223:
            %                           %                    2 -- 10 & \
            %
               \\

            05-01 02:08  & \
            05-01 04:27  & \
            7.9(94\%) & \
            15.4 -- 19.0  \
    %        0  \

            %& Compl.\tablenotemark{b} &
            %               2223:
            %                           %                    16 -- 19 & \
            %
               \\

            05-01 11:51  & \
            05-01 14:10  & \
            7.9(95\%) & \
            13.1 -- 16.0  \
     %       0  \

            %& Compl.\tablenotemark{b} &
            %               2223:
            %                           %                    33 -- 36 & \
            %
               \\

            05-01 17:10  & \
            05-01 19:29  & \
            7.9(94\%) & \
            15.4 -- 19.0  \
     %       0  \

            %& Compl.\tablenotemark{b} &
            %               2223:
            %                           %                    42 -- 45 & \
            %
               \\

            05-03 10:53  & \
            05-05 05:30  & \
            127.3(82\%) & \
            0.6 -- 18.4 \
    %        1  \

            %& Compl.\tablenotemark{b} &
            %               2224:
            %                           %                    4 -- 78 & \
            %
               \\
                                                 \hline
        \end{tabular}
        \end{center}
       	\tablenotetext{a}{The source ON-time takes into account the bad time intervals, due to, e.g., telemetry gaps. In parentheses, the fraction of time with \src\ in the IBIS/ISGRI field of view.}
\end{table}

%&  deg

\section{INTEGRAL data analysis and results} \label{sec:data analysis}

Our results are based mainly on data obtained with the Imager on-board INTEGRAL (IBIS) instrument  \citep{ube03}, that provides images in a 30$\times$30 deg$^2$ field of view thanks to a coded mask coupled to two detector arrays: ISGRI and PICsIT.   ISGRI  collects photon-by-photon events, tagged with a  resolution of 61 $\mu$s, in the nominal 15-1000 keV energy range \citep{leb03}. Higher energies are covered by PICsIT \citep{lab03}, that,  besides other data types,  produces spectral-timing data  with a 7.8-ms integration time allowing  the  detection of impulsive events in the 200-2600 keV energy range.

The position of \src\ was repeatedly in the
field of view of  IBIS, for a total on-time of $\sim$350 ks between 18:54 of April 22 and 05:30 UTC of May 5, 2020  (see  Table~\ref{tab-obslog}).
All the  IBIS/ISGRI data are  analyzed in real time at the INTEGRAL Science Data Center  (ISDC, \citealt{cou03}) by the INTEGRAL Burst Alert System (IBAS), a software for the  automatic search and rapid  localization of gamma-ray bursts and other transient sources \citep{mer03}.

On April 28, IBAS triggered on two bursts, at 09:51:05 UTC and 14:34:24 UTC. It derived their positions with   accuracies  of 3.7 and 2.8 arcmin, respectively,  and distributed the public Alert Packets reporting the identification of \src\ as the origin of these triggers after less than 10 s.
An  analysis of the ISGRI data carried out off-line with the standard imaging analysis software \citep{gol03}
confirmed the association of both events with \src .
For example, the second burst is detected  at coordinates
R.A.=19$^h$ 34$^m$  53.8$^s$,  +21$^{\circ}$ 53$'$ 46$''$  (with 90\% c.l.  error radius   of 1.4$'$), that differ by    0.5$'$   from those of \src .

These two bursts were also detected by the INTEGRAL Spectrometer Instrument\footnote{The signal to noise ratio of the bursts in SPI is much lower  than that in ISGRI  and the data are affected by telemetry saturation features. Therefore,  we do  not  use the SPI data in the present paper.}
(SPI, \citealt{ved03}) and by its Anti Coincidence Shield (ACS, \citealt{von03}).
The  latter provides  light curves with 50 ms resolution at energies above $\sim$80 keV, with maximum sensitivity for directions perpendicular to the satellite pointing  \citep{sav17}.

\subsection{The radio-loud burst}

The brightest of the two bursts discovered by IBAS (burst-G of Table 2) is temporally coincident with  the radio burst  discovered by the CHIME and STARE2  radio  telescopes  \citep{arxiv-CHIME,arxiv-STARE}. Burst-G was also detected   by X-ray instruments on the Insight-HXMT, Konus-WIND and AGILE satellites \citep{arxiv-HXMT,arxiv-konus,arxiv-AGILE}.

\begin{figure*}[ht!]
  \includegraphics[width=18cm]{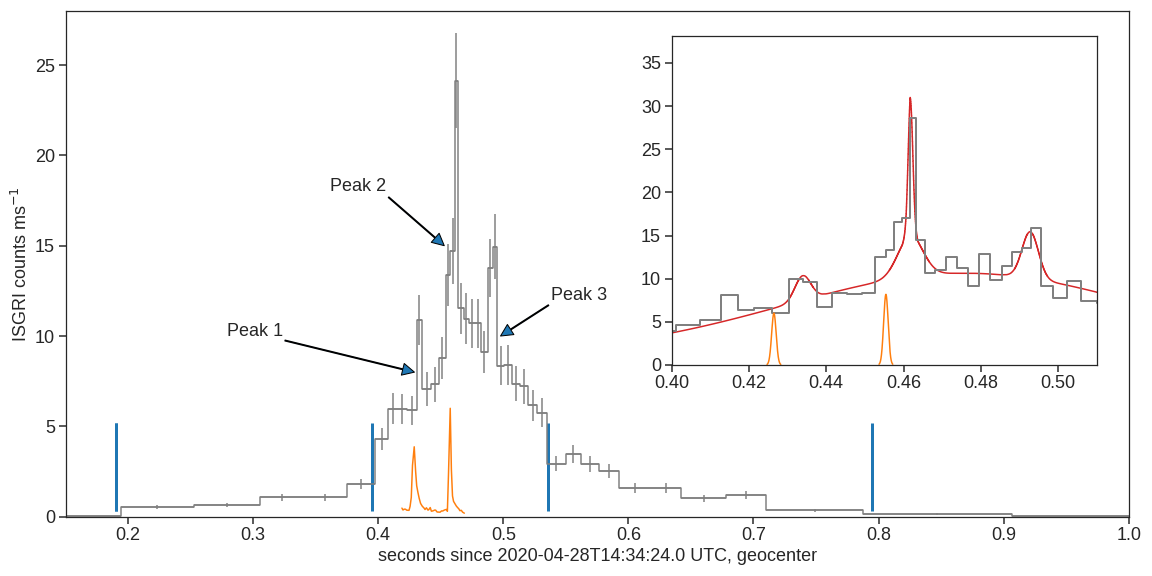}
  \caption{Background subtracted and deadtime-corrected light curve of burst-G in the 20-200 keV range obtained with the IBIS/ISGRI instrument. We used an adaptive binning to ensure at least 40 counts per time bin.   All the times are in the geocentric frame and referred to t$_o$=14:34:24 UTC of April 28, 2020. The blue lines  (at 0.19, 0.395, 0.536, and 0.79 s) indicate the   time intervals used for the spectral and imaging analysis. The orange line (adapted from Fig.~1 of  \citealt{arxiv-CHIME}) marks the position of the radio pulses.
The inset shows the brightest part of the burst, binned  with a minimum of 25 counts per bin.  The red line is the fit to the ISGRI unbinned data with a combination of Gaussian curves (see text for details), that yields the following times for the three narrow peaks:  t$_1$=0.434$_{-0.002}^{+0.004}$~s,  t$_2$=0.462$\pm$0.001 s,    and t$_3$=0.493$_{-0.003}^{+0.002}$~s. The radio pulses are represented in the inset with two Gaussian curves centered at 0.42648 s and 0.45545 s.
 \label{fig:lc}
 }
\end{figure*}

The background-subtracted and dead-time corrected light curve of burst-G, as measured with IBIS/ISGRI in the  20-200 keV energy range  is plotted in Fig.~\ref{fig:lc},
where all the times are in the geocentric frame\footnote{INTEGRAL was  at 128~Mm from Earth, resulting in a delay of 0.195~s for the burst to reach the Earth center. All the times quoted in this paper are corrected for the light travel time from the satellite to the Earth center}.
We  used only detector pixels illuminated by more than 40\% by the source,  and an adaptive binning to get  a minimum of 40 counts per time bin.
We show in  Fig.~\ref{fig:lc} also the position of the CHIME  radio pulses.
The X-ray light curve shows  a main pulse lasting about 0.2 s
with three narrow peaks. To estimate the arrival times of the peaks we fitted the ISGRI unbinned data with a combination of Gaussian curves, taking into account dead time effects while computing the probability of events arrival from the model rate.   The main pulse is well described by two broad Gaussians, and the three peaks by narrow ones centered at  t$_1$=0.434$_{-0.002}^{+0.004}$~s,  t$_2$=0.462$\pm$0.001 s,    and t$_3$=0.493$_{-0.003}^{+0.002}$~s   (we give all the times of burst-G relative to  t$_o$=14:34:24 UTC). The second peak shows a narrow core with two wings, hence we modeled it with the combination of two   Gaussians with the same center.

The main pulse ([0.395--0.536\,s]) is preceded and followed by fainter, slowly varying emission.
Thanks to the ISGRI imaging capabilities, we can verify that such emission is due to the source and not to background fluctuations. In fact, \src\ is clearly detected in the images extracted from the time intervals [0.19--0.395 s]  and [0.536--0.79 s],  with statistical significance of  5.3 and 8.8 $\sigma$, respectively. These intervals are highlighted by vertical blue lines in Fig.~\ref{fig:lc}.

We carried out a spectral analysis using  the Xspec software v. 12.11.0 and
give uncertainties at 68\% confidence level.
To estimate the uncertainties of the best fit parameters we applied
the Monte Carlo Markov Chain method with Goodman-Weare algorithm \citep{Goodman2010} and
derived the $1\sigma$ confidence intervals from the 16 and 84\% percentiles of posteriors.
We used uninformative uniform priors for the slope of the power-law in all models and	uninformative uniform logarithmic priors (known as Jeffreys priors) for all other parameters.
We  ignore channels below 32\,keV, owing to systematic uncertainty in the response calibration and limit our fits below 300\,keV, where the source is significantly detected.
Besides all the effects taken into account by the standard INTEGRAL analysis software, the   fluxes given below are corrected
for the effect of dead-time, which is of 0.1\,ms in each of the eight detector modules that compose the ISGRI instrument. 

We extracted  a total spectrum from a time interval of 0.6 s starting at t$_o$+ 0.19 s,  as well as three spectra for the sub-intervals\footnote{It is possible to extract a spectrum in such a short time
interval with the INTEGRAL Offline Scientific Analysis only starting
from version 11. However, for a source with such a bright burst, we had
to disable the automatic noisy pixel detection, by setting the parameter
\texttt{NoisyDetFlag=0} in the program \texttt{ibis\_science\_analysis}.}
 indicated by the blue lines in Fig.~\ref{fig:lc}.
A single  power-law could not fit well  the total spectrum (photon index $\Gamma=2.1\pm0.1$,  reduced  $\chi_{\nu}^2=3.6$ for 14 degrees of freedom (dof).
Using  an exponentially cutoff power-law, we obtained a best-fit photon index $\Gamma = 0.7_{-0.2}^{+0.4}$, peak energy E$_p = 65\pm5$\,keV, and 20--200 keV flux of $(10.2\pm0.5) \times 10^{-7}\,\mathrm{erg\,s^{-1}\,cm^{-2}}$ ($\chi_{\nu}^2=1.8$ for 13 dof).
 
The X-ray spectra of SGR bursts are often fit also with the sum of two black-body models.
In this case, to avoid that the two components swap in the posterior sampling,  we imposed that one temperature is below 20 keV and the other one above this value in our prior limits.
We found a best fit  with temperatures
$kT_1 = 11.0\pm1.3$\,keV, $kT_2 = 30\pm4$\,keV and radii $R_1= (0.83_{-0.14}^{+0.18})$\,d$_{4.4}$\,km, $R_2=(0.12\pm0.04)$\,d$_{4.4}$\,km ($\chi_{\nu}^2=1.6$ for 12 dof).

To search for possible spectral variations, we first fitted simultaneously the spectra of the three sub-intervals with the cut-off power law model, keeping both $\Gamma$ and $E_p$ tied among the three spectra,
and then checked with the F-test statistics the significance of fit improvements obtained with either or both parameters free to vary.
Imposing only a common slope, we obtained
$\chi^2$/dof=46/41 (compared to $\chi^2$/dof=58/43 with both parameters tied), that corresponds to  0.7\% chance probability of the  fit improvement. 
The best fit parameters are $\Gamma=0.62_{-0.18}^{+0.22}$,   E$_p$  of  $34\pm8$  keV (before main pulse),  $60\pm5$  keV (main pulse), and $125_{-29}^{+50}$\,keV (after main pulse), and 20--200\,keV fluxes of
$1.9_{-0.3}^{+0.2}$, $32^{+2}_{-1}$, and $4.6_{-0.6}^{+0.4}$ in units of $10^{-7}\mathrm{erg\,s^{-1}\,cm^{-2}}$, respectively.
This result gives some evidence for a spectral hardening as a function of time.
Keeping instead a common $E_p$ and letting the slopes as free parameters gave a worse fit  ($\chi^2$/dof=52/41), indicating that the spectral evolution is driven by an increase of the peak energy, rather than a change in slope.
By leaving all parameters free, we found  no statistically significant improvement with respect to using a common slope.

We extracted also the spectra of the three narrow peaks, but their small number of counts does not allow  a meaningful spectral analysis. Adopting the best fit cut-off power law parameters of the total spectrum, we obtained the following 20--200 keV fluxes in units of 10$^{-7}$ erg cm$^{-2}$ s$^{-1}$:   36$_{-17}^{+7}$  (Peak-1, [0.431--0.435 s]),   58$_{-8}^{+5}$  (Peak-2, [0.454--0.464 s]) , and  54$_{-8}^{+6}$  (Peak-3, [0.487--0.495 s]).

Analysis of the PICsIT spectral-timing data corresponding to the total time interval  yielded a 3$\sigma$ upper limit of 1.6$\times10^{-7}$  erg  cm$^{-2}$  s$^{-1}$
in the 200--500 keV energy range, assuming the cut-off power law spectral parameters from ISGRI.

The SPI/ACS data are routinely analyzed to search for bursts on different time scales. Burst-G reached the highest signal to noise ratio (4.65) in the data binned at 0.65 s.  We note that, based only on time coincidence (following \citealt{con16}) and independently of the IBIS/ISGRI detection, the ACS data give an association with the radio burst with a false alarm probability at the level of 2.9 $\sigma$.

\subsection{Other bursts}

A more sensitive  search  for other bursts in the IBIS/ISGRI data was carried out off-line, by examination of mask-tagged light curves of \src\ on timescales of 10~ms, 100~ms, and 1~s, extracted using the \texttt{ii\_light} program.
This resulted in the discovery of other seven bursts using the 100 ms timescale (see Table \ref{tab-burst}),  that were missed by the IBAS software  due to their large off-axis angles.

\begin{figure}[ht!]
\centering
\includegraphics[width=8cm]{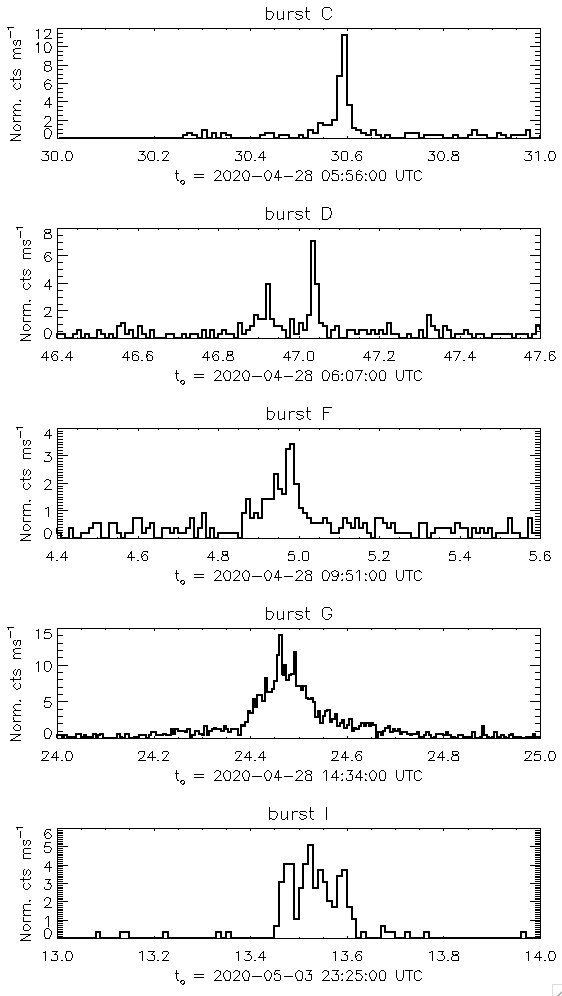}\caption{Light curves in the 20-200 keV range for the bursts of Table 2 with more than 100 counts.
Time bins are of 10 ms, except  for burst-G that is binned at 5 ms. The count rates have been normalized to the values that would have been obtained with the source on-axis.
\label{fig:lc-other}}
\end{figure}

We performed a spectral analysis for the bursts with more than 100 counts  (bursts C, D, F, and I; their light curves are shown in Fig.~\ref{fig:lc-other}). In all cases a power-law model gave an acceptable fit  and the data could not constrain the parameters of more complex models.  We then computed the weighted average of their photon indexes, finding $\bar \Gamma =3.2\pm0.5$ and  imposed a Gaussian prior with these values to the photon index in the fits of the other bursts. We give all the results in  Table \ref{tab-burst}, where,  for the sake of comparison, also the parameters of burst-G refer to the power law fit.
All the bursts detected with high signal to noise ratio have  spectra   softer than that of burst-G.

\begin{figure}
	\includegraphics[width=\columnwidth]{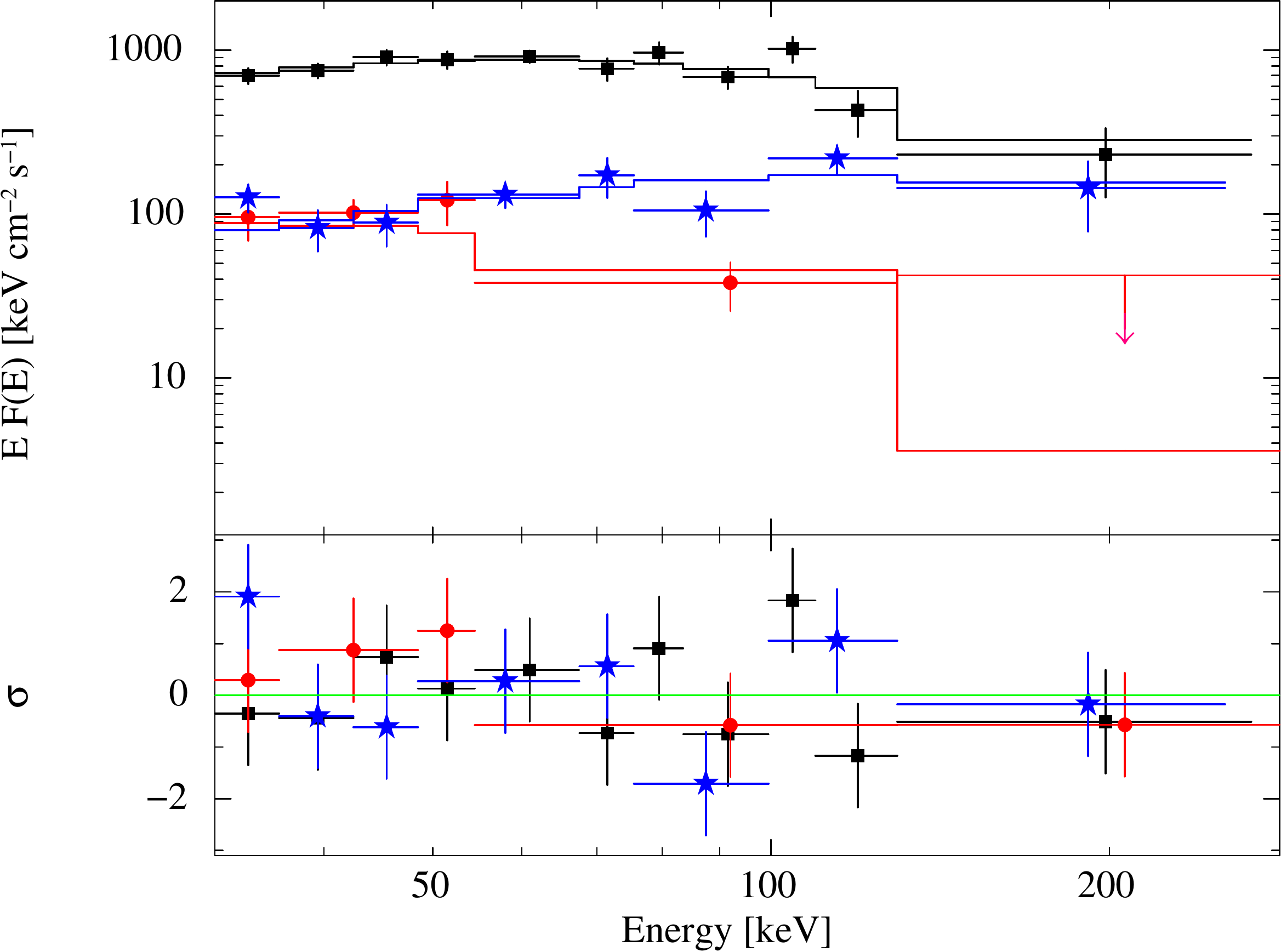}
	\label{fig:burst_spectra}
	\caption{Unfolded energy spectra of burst-G fitted with a cut-off power law model (top panel). The three spectra refer to the main  pulse (black squares), and to the intervals before (red circles) and after (blue stars) the main pulse. The spectra have been rebinned for plotting purposes. The lower panel shows the residuals in units of $\sigma$.
	\label{fig:sp3}
	}
\end{figure}

\begin{table*}
    \caption{Bursts observed in the selected observation period (see Table~\ref{tab-obslog}).}
    \begin{center}
    \begin{tabular}{lccccccc}
        \hline
        Name & Trigger Time       &  Duration                 & S/N\tablenotemark{a} &  Total\tablenotemark{b} & Fluence       & $\Gamma$  & off-axis \\ 
        & {\scriptsize YYYY-MM-DD HH:MM:SS.S}  &    s                      &                      &  counts  & $10^{-7}$erg\,cm$^{-2}$ &   & angle($^o$)  \\
        \hline
        \hline

            A & \
            2020-04-28 03:47:52.200 & \
            0.06 & \
            17 & \
            33 (2.1) & \

				            	              1.3\small$^{+0.2}_{-0.3}$\normalsize& \

				       		-- & \
	        	            
            15.2 \\
		                    
            B & \
            2020-04-28 04:09:47.300 & \
            0.09 & \
            14 & \
            35 (2.7) & \

				            	              1.0\small$^{+0.2}_{-0.3}$\normalsize& \

				       		-- & \
	        	            
            15.2 \\
		                    
            C & \
            2020-04-28 05:56:30.637 & \
            0.11 & \
            14 & \
            113 (10.2) & \

				            	              1.4\small$^{+0.2}_{-0.1}$\normalsize& \

	                	            3.0\small$^{+0.5}_{-0.4}$\normalsize& \
        	    	        	            
            12.4 \\
		                    
            D & \
            2020-04-28 06:07:47.037 & \
            0.17 & \
            12 & \
            110 (18.1) & \

				            	              1.0\small$^{+0.2}_{-0.1}$\normalsize& \

	                	            3.2\small$^{+0.6}_{-0.3}$\normalsize& \
        	    	        	            
            12.4 \\
		                    
            E & \
            2020-04-28 08:03:34.370 & \
            0.04 & \
            9 & \
            18 (1.8) & \

				        	$<$ 0.7 & \

				       		-- & \
	        	            
            14.2 \\
		                    
            F & \
            2020-04-28 09:51:04.894 & \
            0.38 & \
            10 & \
            201 (56.0) & \

				            	              1.0\small$^{+0.4}_{-0.1}$\normalsize& \

	                	            3.8\small$^{+1.2}_{-0.4}$\normalsize& \
        	    	        	            
            10.4 \\
		                    
            G\tablenotemark{c} & \
            2020-04-28 14:34:24.357 & \
            0.75 & \
            23 & \
            1629 (159.1) & \

				            	              6.9\small$^{+0.3}_{-0.2}$\normalsize& \

	                	            2.1\small$^{+0.1}_{-0.1}$\normalsize& \
        	    	        	            
            8.1 \\
		                    
            H & \
            2020-04-29 09:10:53.895 & \
            0.06 & \
            8 & \
            7 (0.7) & \

				        	$<$ 0.8 & \

				       		-- & \
	        	            
            15.6 \\
		                    
            I & \
            2020-05-03 23:25:13.469 & \
            0.15 & \
            46 & \
            136 (3.1) & \

				            	              10.5\small$^{+1.4}_{-1.2}$\normalsize& \

	                	            3.3\small$^{+0.5}_{-0.3}$\normalsize& \
        	    	        	            
            16.9 \\
		         
        \hline
    \end{tabular}
	\end{center}
	
	\tablenotetext{a}{Signal to noise ratio extracted from the lightcurve binned at 100\,ms  used to detect the bursts.}
	 \tablenotetext{b}{Total raw counts collected by IBIS in the indicated time range; in parentheses the number of counts expected from the background.}
	 \tablenotetext{c}{This is the burst with associated radio emission. The spectral parameters given here are those of the power law fit, to facilitate the comparison with the other bursts. See the text for the more appropriate values obtained with other spectral models.}
 
	 \label{tab-burst}
    
\end{table*}

\begin{figure*}[ht!]
\centering
\includegraphics[width=12cm]{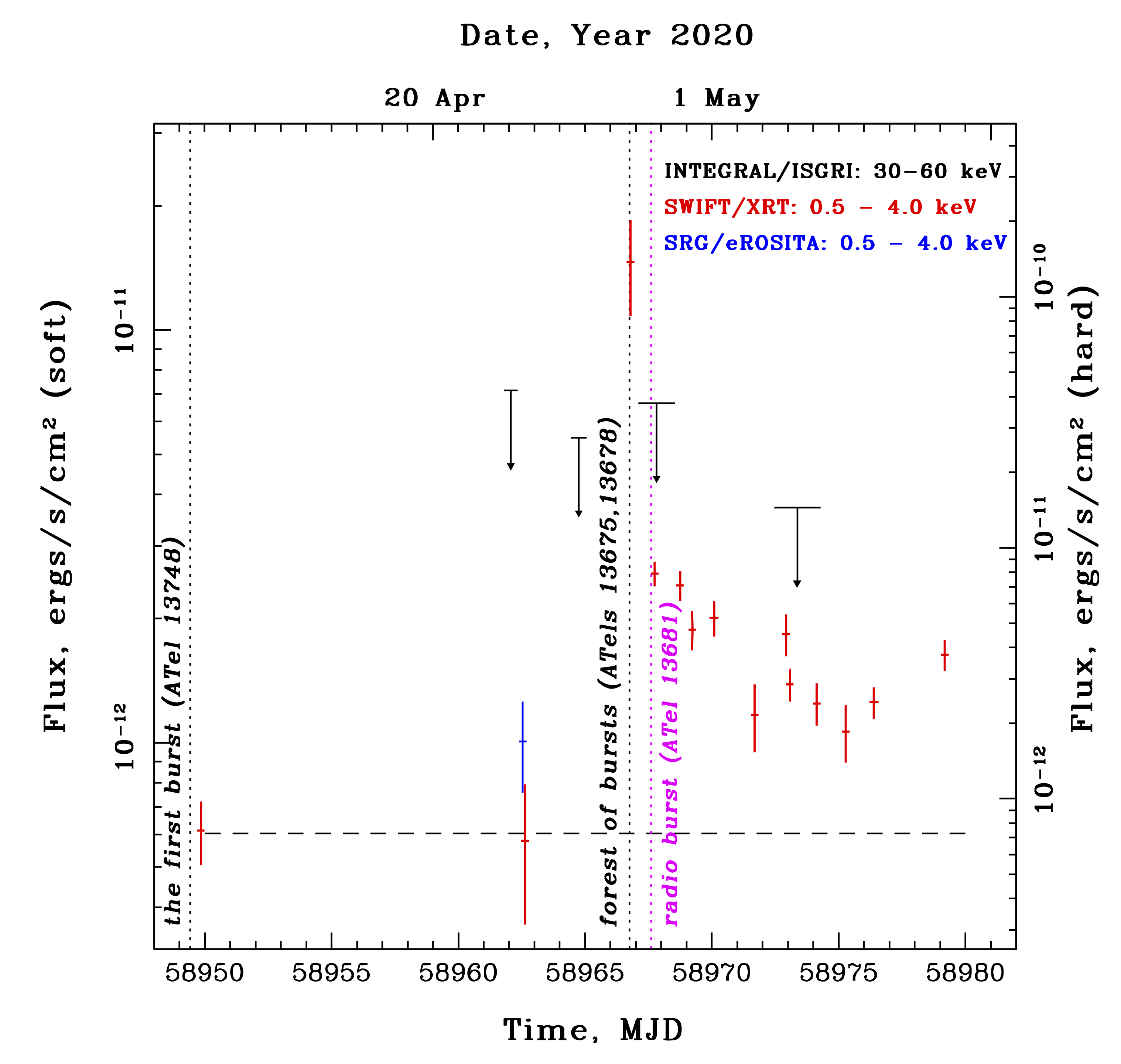}\caption{Evolution of the \src\ mean flux measured in soft X-rays
with Swift/XRT (red points) and SRG/eROSITA (blue point) and upper limits
from INTEGRAL in hard X-rays (black points). The horizontal dashed line
shows the persistent flux level in the 0.5-4 keV energy band, calculated with
parameters from \citet{you17}. See text for details.
\label{fig:longterm}}
\end{figure*}

A highly polarized  burst from \src\ was detected by  the FAST radiotelescope on April, 30 at 21:43:00.5 UTC \citep{atel13699}. At this time, \src\  was in the field of view of IBIS at an off-axis angle of 12.9$^o$, and we could derive a 5$\sigma$ upper limit on the fluence of any burst shorter than 1~s at the level of 2.3$\times10^{-8}\,\mathrm{erg\,cm^{-2}}$ in the 20--200\,keV band, assuming the spectral shape of burst-G. This corresponds to a ratio of radio to X-ray fluence at least six orders of magnitude smaller than that of burst-G.

 \subsection{Persistent emission}

No persistent hard X-ray  emission  from \src\ was   detected by IBIS/ISGRI.
We show in   Fig.~\ref{fig:longterm} the upper limits    derived on its 30-60 keV flux from the four longest INTEGRAL observations (the first three and the last one of Table 1).
To trace the overall behaviour of the source, we show in the same figure its lightcurve obtained
from publicly available data of the Swift/XRT telescope,  covering the period
from April 10 to May 10, and the  SRG/eROSITA observations performed during the
all sky survey on April 23 \citep{atel13723}.
The light curve in the 0.5-4 keV energy band  demonstrates that the
soft X-rays from \src\ significantly increased more than 20 times
in less than four days before the start of the bursting activity
and began to subside practically immediately after its end (see also \citealt{arxiv-bor20})

The expected source flux in hard X-rays can be estimated using the
spectral parameters obtained with NuSTAR in the 3-79 keV energy range
\citep{you17}. Assuming that the spectral shape does not evolve
with the flux and renormalizing the spectra with the measured fluxes
in the soft energy band, we found that all IBIS/ISGRI upper limits are above the
expected  hard X-rays fluxes. They are also consistent with
the source flux of 4$\times10^{-12}$\,erg\,cm$^{-2}$\,s$^{-1}$ derived from the NuSTAR
observation performed on May 2 \citep{arxiv-bor20}.

\begin{figure*}[ht!]
\centering
\includegraphics[width=15cm]{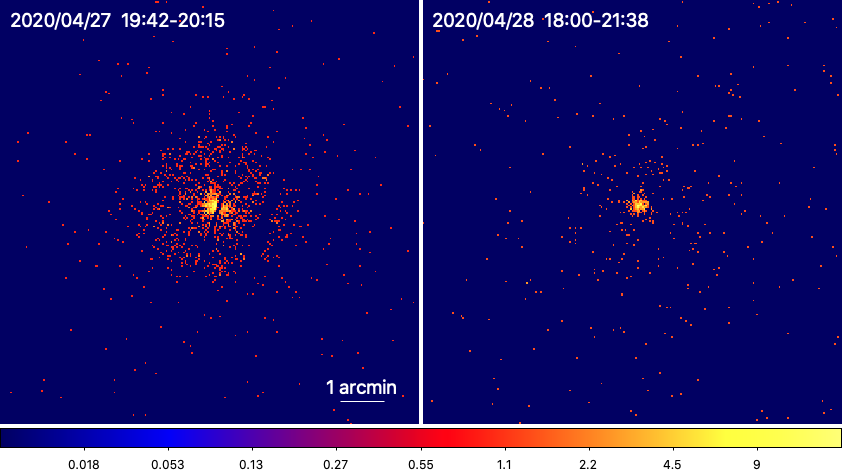}
\caption{X-ray images (0.2-10 keV) of \src\ obtained with  Swift/XRT on 2020 April 27 (ID 00968211001, left panel) and  28 (ID 00033349045, right panel).  The first observation clearly shows a ring of scattered  emission  that extends for more than 1$'$ from the source. The ring was no more visible the following day, owing to its fading X-ray flux and increasing angular radius.
\label{fig:ring}}
\end{figure*}

\section{Distance estimate from the dust scattering ring}\label{sec:dist}

A bright X-ray ring, with radius of $\sim$85$''$, was detected around \src\ in a  Swift/XRT observation performed on April 27, 2020 from 19:42 to 20:15 \citep{atel13679}. Its very rapid evolution (see Fig.~\ref{fig:ring}) excludes the possibility of a spatially extended structure at any plausible distance for \src\ (e.g., a magnetar wind nebula) and resembles that of   rings produced by scattering of X-rays on interstellar dust grains, as seen around some GRBs and after extremely bright magnetar bursts (see, e.g., \citealt{tie10}).
Expanding scattering rings produced by bursting sources give the possibility to derive some information on the   distances of the emitting source and of the scattering dust by studying the time delay of scattered photons \citep{tru73}.

We extracted from the XRT data (obs.ID 00968211001) three time-resolved radial profiles in the 1--5\,keV energy band by dividing the observation into 11 minutes intervals. The  profiles were well fit  with the sum of a Lorentzian, a King and a constant function, to model the ring,  the \src\ unscattered emission and the background, respectively. The centers of the Lorentzian curves found in the three time intervals, 79.4$\pm$1.9, 85.2$\pm$2.5 and 93.2$\pm$4.4 arcsec,  clearly indicate that the ring significantly expanded during this short observation.

Due to the time delay of scattered photons, the angular radius $\theta$ of a ring produced by a burst emitted at $t_B$ by an X-ray source at distance $d_s$ and scattered by dust at distance $d_d$, increases with time $t$ as:
\begin{equation}
\theta (t) = \bigg[\frac{827}{d_s} \frac{(1-x)}{x} (t-t_B)\bigg]^{1/2}
\label{ringexp}
\end{equation}
where the distances are measured in parsecs,  $\theta$  in arcseconds, times in seconds, and $x=\frac{d_d}{d_s}$.

Fitting the ring expansion with eq.~\ref{ringexp}, we found  $t_B$  in the 68\% c.l. interval from 18:00 to 18:50 of  April 27.
The main bursting activity of \src\ started on April 27 at 18:26:20  \citep{atel13675}, but the largest X-ray fluence was produced about 5 minutes later, in a series of bright bursts emitted within $\sim$10 s \citep{atel13682}.  No other bursting episodes with similar fluence were detected in the preceding and following hours \citep{tav20}. It is therefore reasonable to attribute the origin of the X-ray ring to  this ``burst forest'', and fix $t_B$ at 18:33:10.

Based on eq.~\ref{ringexp}, using the method of \citet{tie06}, we   assigned a pseudo-distance $D_i=827 (t_i-t_B)\theta^{-2}$ to each event $i$ detected by XRT.  The distribution of $D_i$ values shows a clear peak centered at $D$=580$\pm$10 pc. For an extra-galactic source scattered by Galactic dust ($d_s>>d_d$), as, e.g., in \citet{tie06} and \citet{pin17}, $D$ would be a very good approximation of the dust layer distance producing the X-ray ring, whereas in the general case $D=d_d/(1-x)$.

The distribution of interstellar dust  towards \src\ is well characterized in the 3D extinction maps of \citet{gre19}, based on Pan-STARRS 1 and 2MASS photometry. and on Gaia parallaxes. These maps show that most of the optical extinction is caused by a narrow dust layer at $\sim$0.5 kpc, while the rest is due dust concentrations at $\sim$1.1 and $\sim$6.5 kpc. The  $D$ value derived above gives a hard upper limit on $d_d$, implying that  only the closest dust layer can be responsible for the X-ray ring. To derive  $d_d$ more accurately, we used the Python package $dustmaps$ \citep{gre18} to extract the differential extinction in the \src\ direction from 400 to 1000 pc, with a 5 pc step, by taking the median of the 5 samples available in the \citet{gre19} map, weighted on their standard deviations. The resulting extinction distribution can be well fit by a single Gaussian peak centered at 512 pc, with a standard deviation of 16 pc
\footnote{To validate this method for measuring $d_d$, we applied the same procedure to the line of sight of GRB 160623A, where six dust-scattering rings were detected with \xmm\  \citep{pin17}. We obtain a good fit of the extinction differential distribution in the 400--1000 pc interval  using the sum of three Gaussians centered at 519, 660  and 786 pc, and  with standard deviations of 11, 36 and 3 pc, respectively. These distances are in very good agreement with those of the three closest dust layers derived independently (the GRB can be considered at infinite distance) from the analysis of the dust scattering rings: 528.1$\pm$1.2 pc, 679.2$\pm$1.9 pc and 789.0$\pm$2.8 pc}. Using this value for the dust distance, we
obtain a distance of  $d_s=4.4_{-1.3}^{+2.8}$ kpc  for \src , independent of, but consistent with, its association with the supernova remnant G57.2+0.8.

\begin{figure}[ht!]
\centering
\includegraphics[angle=270,width=10cm]{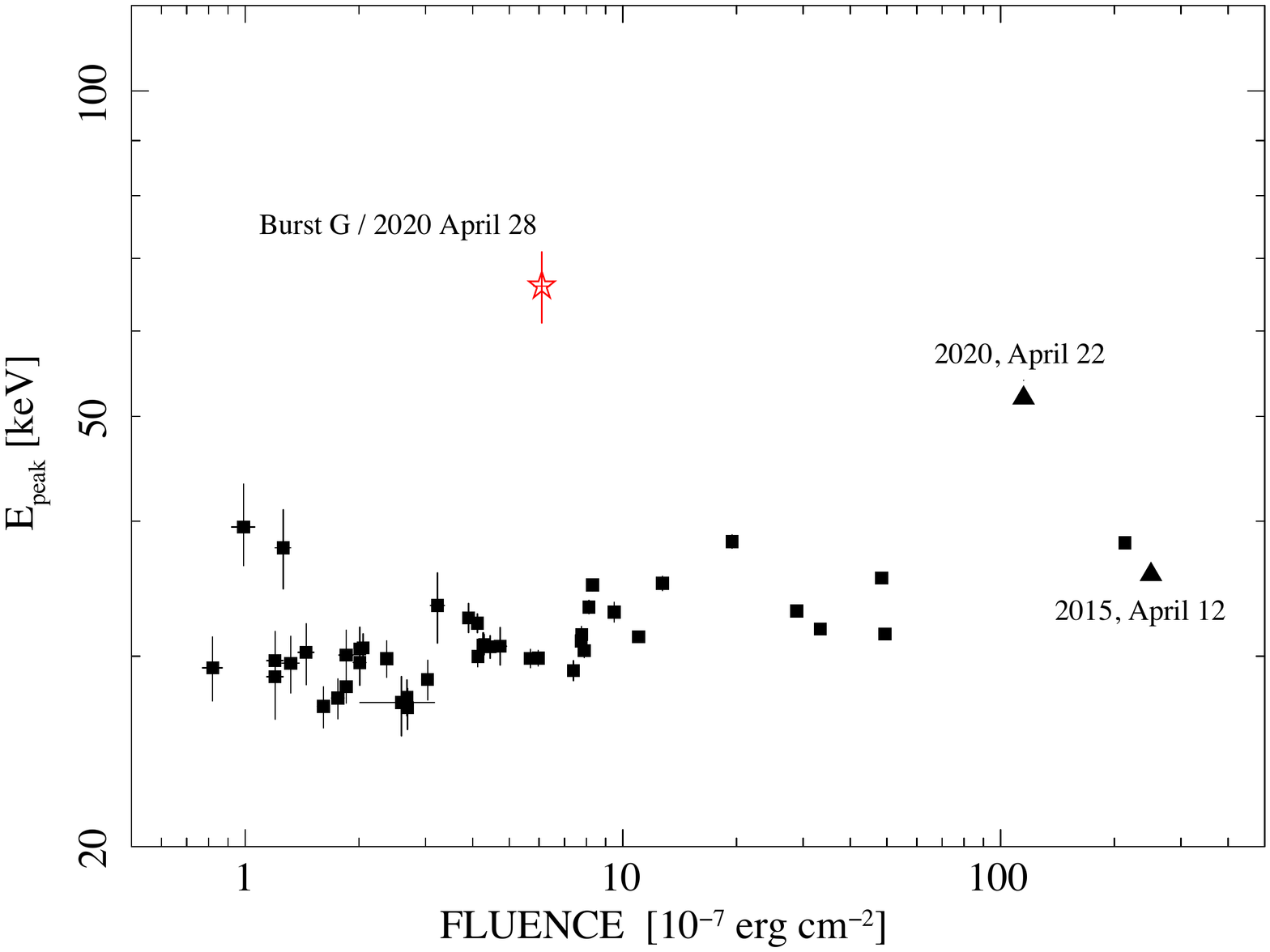}
\caption{Peak energy versus fluence for Burst-G (red star) and other bursts from \src\  (squares from \citet{lin20}, triangles from \citet{koz16} and \citet{gcn27631}.
\label{fig:flukt}}
\end{figure}

\begin{figure*}[ht!]
\centering
\includegraphics[width=2.\columnwidth]{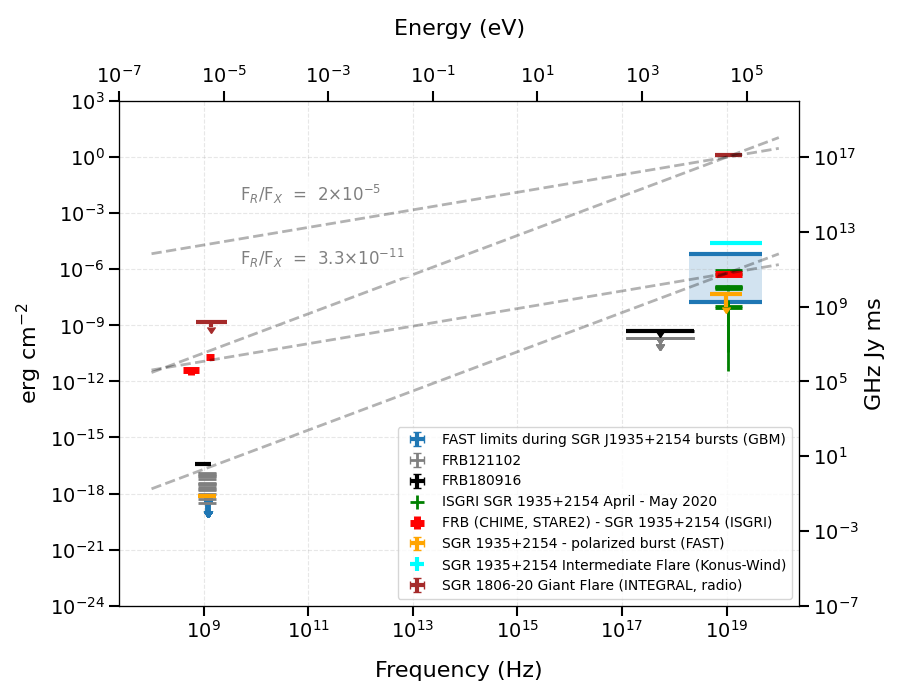}
\caption{ Comparison of the spectral energy distribution of burst-G from \src\ (red) with the upper limits for other magnetars and FRBs obtained from
 simultaneous   radio and X-ray observations \citep{hur05,ten16,koz16,sch17,arxiv-FAST, arxiv-CHIME, arxiv-STARE,sch20,pil20}
\label{fig:sed}}
\end{figure*}

\begin{figure*}[ht!]
\centering
\includegraphics[width=1.8\columnwidth]{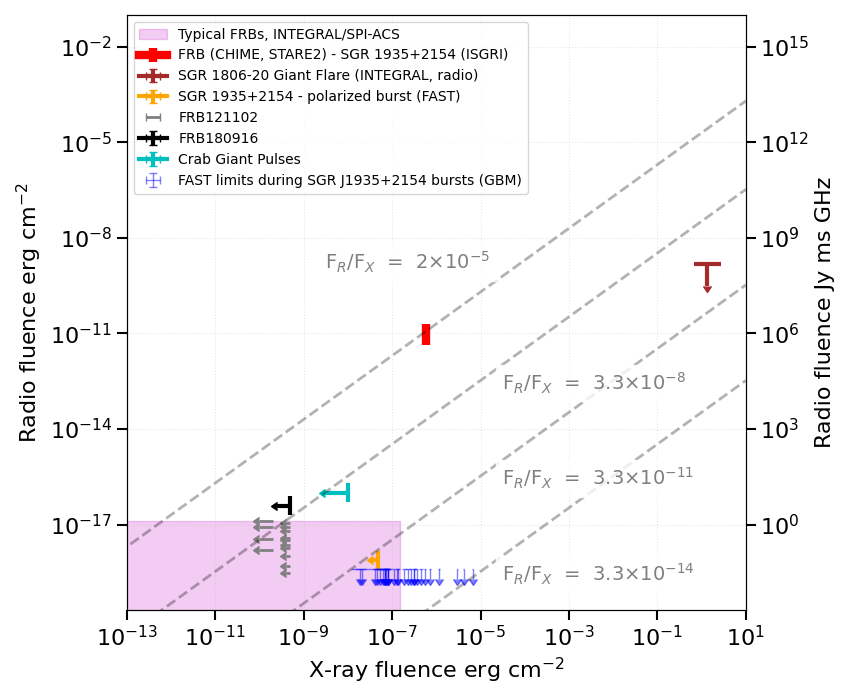}
\caption{ \label{fig:FxFr} Radio versus  X-ray (starting from 0.5~keV)  fluences  for FRBs and magnetar bursts.
The range of FRB fluences corresponds to a variety of detections reported in the past years (references in Fig.~\ref{fig:sed} and \citealt{kar10}). The purple region indicates a robust upper limit on the hard X-ray fluence of FRBs as derived with a high-duty-cycle detector, such as the INTEGRAL SPI/ACS \citep{sav17}.
}
\end{figure*}

\section{Discussion} \label{sec:discussion}

Burst-G is the first burst from an SGR to be  unambiguously associated with a short pulse of radio emission with some properties similar to those of FRBs
%that, apart from a DM   not exceeding the Galactic value and a much lower luminosity, shows characteristics very similar  to  those of extragalactic FRBs 
\citep{arxiv-CHIME,arxiv-STARE}.
%\citep{atel13681,atel13684}.
It is, therefore,   interesting to investigate if its high-energy properties  differ from those of the bursts commonly emitted by this and other magnetars.

The 20-200 keV   fluence of $(6.1\pm0.3) \times 10^{-7}\,\mathrm{erg\,cm^{-2}}$
and peak flux of  $\sim6 \times 10^{-6}\,\mathrm{erg\,s^{-1}\,cm^{-2}}$,
derived from the IBIS/ISGRI  spectrum with the cut-off power law model,
are in the range of values previously observed for other  bursts of \src\ \citep{lin20}.
Therefore, they do not qualify burst-G  as particularly energetic.
They correspond to a  peak luminosity of    $\sim10^{40} d_{4.4}^2$ erg s$^{-1}$    and to a released energy of $\sim1.4 \times 10^{39} d_{4.4}^2$ erg  in the 20-200 keV range (assuming isotropic emission).

On the other hand, as suggested by the comparison with the other bursts in Table~\ref{tab-burst}, the spectrum of burst-G was harder than those of  typical SGR bursts.
This can be seen, for example, by comparing its peak energy E$_p$=65$\pm$5 keV, with the values derived  by
 \citet{lin20} for  a large sample of bursts from \src\ observed with the Fermi/GBM instrument (see Fig.~\ref{fig:flukt}).
The distribution of E$_p$ found by these authors is well fitted by a Gaussian centered at 30.4  keV and  with $\sigma$= 2.5   keV. A similar conclusion is obtained if we compare the results of the two blackbody fit, for which we obtained temperatures   a factor $\sim2.5$ larger than the average values of kT$_1$=4.4  keV and  kT$_2$= 11.3 keV   found by \citet{lin20}. 
From the spectral point of view, the \src\ burst most similar to burst-G was the one detected by several satellites on April 22, 2020 \citep{gcn27623,gcn27625}. It  had    a peak energy of E$_p$=52$\pm$2 keV and a 20-200 keV fluence of $\sim10^{-5}$ erg cm$^{-2}$  (Fig.~\ref{fig:flukt}),  but a  simple time profile consisting of a single pulse lasting 0.6 s with   fast rise  and no evidence for sub-structures \citep{gcn27631}.
Notably, the spectrum of burst-G was also harder than that of the intermediate flare emitted by \src\   on
April 12, 2015.
This was at least a factor 10 more energetic  (Fig.~\ref{fig:flukt}),  but  had E$_p$=36 keV and a smooth time profile \citep{koz16}.

It is thus possible that the radio emission, detected so far only in burst-G, might be related to its particularly hard X-ray spectrum. \citet{arxiv-konus} noted that another distinctive property of burst-G might be its spiky and slowly rising light curve. The small number of counts of the other bursts we detected does not allow us to draw robust conclusions on this  interesting possibility (Fig.~\ref{fig:lc-other}). 

The triple peaked time profile of burst-G shown in Fig.~\ref{fig:lc}  is particularly interesting  because the time intervals between the X-ray peaks (28$_{-4}^{+2}$ ms and 31$_{-3}^{+2}$ ms) are consistent with the 29 ms separation of the two radio pulses seen by CHIME \citep{arxiv-CHIME}. The narrower component of Peak 2 is centered 6.5$\pm$1.0 ms after the second CHIME pulse (that is at the same time of the STARE2 pulse, \citealt{arxiv-STARE}), but its broader component starts at a time consistent with that of the radio pulse. A similar delay, but with a larger uncertainty, is found between peak 1 and the first CHIME pulse. A physical association of the radio pulses with peaks 2 and 3 would instead imply a delay of $\sim$35 ms between the X-ray and radio emission. Note that the accuracy in the relative time alignment of the radio and X-ray light curves is limited by the small number of X-ray counts and not by the absolute time error of the INTEGRAL data that is smaller than 0.1 ms \citep{Kuiper2003}.

The close time coincidence of the radio and X-ray emission is consistent with an origin of both components in a relatively small region of the pulsar magnetosphere (e.g., \citet{lyut02,wad19,lyub20}).  On the other hand, also models involving emission at distances much larger than the light cylinder radius ($Pc/2\pi = 1.5\times10^{10}$ cm for \src ) can produce (nearly) simultaneous pulses due to relativistic Doppler effects  (e.g., \citep{bel19,mar20a,mar20b}).

Explanations involving  magnetars at extragalactic or cosmological distances have been among the first ones to be  proposed for FRBs  \citep{pop07}.  Several models have been developed, based on scenarios for which high-energy emission   can be expected in the form of prompt bursts (e.g. \citet{lyub14,bel17,bel19,lyub20}) or as a long lasting afterglow following the FRB \citep{mur17}. However, all the  searches for X/$\gamma$-ray counterparts of FRBs carried out up to now have been unsuccessful\footnote{The claimed association of a $\sim$400 s long  X-ray transient with  FRB~131104  \citep{del16} has not been supported by subsequent observations \citep{sha17}. In any case, the  properties of this transient are very different from those of burst-G.} \citep{sch17,xi17,mar19,gui19,gui20,cun19,sun19,anu20}.

\citet{ten16} derived lower limits on the ratio between radio (1.4 GHz)  and X/$\gamma$-ray ($>$30 keV) fluences $\eta_{FRB}>10^{7-9}$ Jy ms erg$^{-1}$ cm$^2$ for 15 FRBs.
Negative results were found  also with coordinated multiwavelength observations of   \frbdic\ \citep{sch20,pil20,tav20}, that has been considered as a promising target, owing to its close distance of 150 Mpc and periodic behavior of the radio burst emission \citep{CHI20}.
Similarly,   searches for FRBs associated with magnetar  bursts provided only upper limits \citep{bur18}, including the case of the December 2004 giant flare from  SGR 1806$-$20, for which \citet{ten16} derived  $\eta_{FRB}<10^7$ Jy ms erg$^{-1}$ cm$^2$.

Based on the radio fluences of  700 kJy ms at 600 MHz \citep{arxiv-CHIME} and 1.5 MJy ms at 1.4 GHz  \citep{arxiv-STARE}, 
%    reported for FRB 200428,
we  compare the  spectral energy distribution of burst-G  (Fig.~\ref{fig:sed}) and the ratio of its radio and X-ray fluences F$_R$/F$_X$  (Fig.~\ref{fig:FxFr}) with the  limits obtained for other relevant events.

Burst-G was characterized by a very large value of F$_R$/F$_X$  $\sim2\times10^{-5}$.
This high ratio is consistent with the lack of detection of high-energy emission in extragalactic FRBs, as it can be seen by scaling the fluences to distances of a few hundreds of Mpc. On the other hand, such a high F$_R$/F$_X$ is at variance with the limit derived for the SGR 1806$-$20 giant flare. This   suggests that giant flares and ordinary SGR bursts are indeed different phenomena, as it is also supported by their different spectral and timing properties,  and/or that magnetar bursts can have a large range of F$_R$/F$_X$, as shown by the radio upper limits reported by \citet{arxiv-CHIME} and \citet{arxiv-FAST}.
It must also be considered  that both the radio and high-energy emissions could be non-isotropic and beamed in different directions, thus complicating the observational picture and making it difficult to draw strong conclusions until a much larger sample of bursts is observed.

\section{Conclusions} \label{sect:conclusions}

INTEGRAL has discovered a peculiar hard X-ray  burst, characterized by  the simultaneous emission of a very bright millisecond radio pulse \citep{arxiv-CHIME,arxiv-STARE},  from the Galactic magnetar \src .
This burst was not particularly energetic at high energies, but it differed from the typical SGR bursts because of its harder spectrum.
The discovery of simultaneous fast bursting emission at radio and high-energy from this Galactic source supports models based on magnetars for  extragalactic FRBs, although the latter involve a   larger  radio energy output compared to the case discussed here.

With an analysis of the dust scattering X-ray ring observed  with Swift/XRT about one hour after a period of particularly intense bursting emission, we derived a distance of 4.4$_{-1.3}^{+2.8}$ kpc for \src .  We note that this value, contrary to other distance estimates, does not depend on the  association of the magnetar with the supernova remnant G57.2+0.8.

If all extragalactic FRBs were characterized by ratios of radio to X-ray fluences as large as that of burst-G from \src , 
their detection with current high-energy satellites would be difficult. However, since it is clear that a single fluence  ratio is not compatible with all available measurements (see~Fig.~\ref{fig:FxFr}),   future multi-wavelength (and especially high-duty-cycle) observations  will reveal how unusual was the, so far unique, joint detection obtained for \src\ and possibly confirm if, 
as it is suggested by the INTEGRAL data, radio emission is associated only to the spectrally-hardest SGR bursts or it is a more common property.

Finally, we would like to make a comment on the role of magnetars in the context of both Gamma-ray bursts (GRBs) and FRBs. In the first years following their discovery, GRBs and SGRs were believed to belong to the same phenomenological class of short-lived high-energy transients of mysterious origin. However, SGRs’ distinguishing features of repetitive behaviour, giant flares, pulsations and  association with supernova remnants along the galactic plane, convincingly demonstrated they belonged to a different population, that of galactic magnetars. On the other hand, GRBs were confirmed as cosmological sources, putting to rest one of the biggest controversies in astrophysics of the previous few decades. Newly born magnetars can also be considered as  possible central engines for powering extragalactic GRBs, while a small fraction of the short-duration GRBs may be due to extragalactic magnetar giant flares.

Mirroring this GRB-SGR-magnetar entanglement, the observation of FRB-like radio emission and gamma-ray flares from the known galactic magnetar \src\ now opens the possibility that  some of the sources that have been classified as FRBs
consist of galactic magnetars so far unidentified at other wavelengths, while also providing strong support for a magnetar origin of extragalactic FRBs.

\acknowledgments
Based on observations with INTEGRAL, an ESA project with instruments and science data centre funded by ESA member states (especially the PI countries: Denmark, France, Germany, Italy, Switzerland, Spain) and with the participation of Russia and the USA.
ISGRI has been realized and maintained inflight by CEA-Saclay/Irfu with the support of CNES.
The Italian authors acknowledge  support via ASI/INAF Agreement n. 2019-35-HH and PRIN-MIUR 2017 UnIAM (Unifying Isolated and Accreting Magnetars, PI S.Mereghetti).
D.G. acknowledges the financial support of the UnivEarthS LabEx (ANR-10-LABX-0023 and ANR-18-IDEX-0001).
A.L. and S.M. acknowledge the financial support of the Russian Foundation of Basic Research (proj. 19-29-11029).
L.H. acknowledges support from Science Foundation Ireland grant 09/RF/AST2400.
We are grateful to the Swift/XRT team for making  observations open for the broad community.



\bibliographystyle{aasjournal} % style aasjournal.bst
\bibliography{1935}

%\begin{thebibliography}{}
%
%\end{thebibliography}



%\appendix

%\input{}

\end{document}